\documentclass[twocolumn,aps,showpacs]{revtex4}
\usepackage{amssymb}
\usepackage{amsmath,bm}
\usepackage{graphicx}
\usepackage[normalem]{ulem}
\usepackage{multirow}

\usepackage[usenames, dvipsnames]{xcolor}
\usepackage{lipsum}

\renewcommand{\sout}{\bgroup \color{red} \ULdepth=-.5ex \ULset}

\begin{document}

\title{Suppression of light  nuclei production in collisions of small systems at the Large Hadron Collider}

\author{Kai-Jia Sun\footnote{%
email: kjsun$@$tamu.edu}}
\affiliation{Cyclotron Institute and Department of Physics and Astronomy, Texas A\&M University, College Station, Texas 77843, USA}
\author{Benjamin D\"{o}nigus\footnote{%
b.doenigus@gsi.de}}
\affiliation{Institut f\"{u}r Kernphysik, Johann Wolfgang Goethe-Universit\"{a}t Frankfurt, Max-von-Laue-Str. 1, 60438 Frankfurt, Germany}
\author{Che Ming Ko\footnote{%
ko@comp.tamu.edu}}
\affiliation{Cyclotron Institute and Department of Physics and Astronomy, Texas A\&M University, College Station, Texas 77843, USA}

\date{\today}

\begin{abstract}
We show that the recently observed suppression of the yield ratio   of deuteron to proton and  of helium-3 to proton in p+p collisions compared to those in p+Pb or Pb+Pb collisions by the ALICE Collaboration at the Large Hadron Collider (LHC) can be explained if light nuclei are produced from the coalescence of nucleons at the kinetic freeze-out of these collisions. This suppression is attributed to  the non-negligible sizes of deuteron and helium-3 compared to the size of the nucleon emission source in collisions of small systems, which reduces the overlap of their internal wave functions with those of nucleons.  The same model is also used to study the production of triton and hypertriton in heavy-ion collisions at the LHC. Compared to helium-3 in   events of low charged particle multiplicity, the triton is less suppressed due to its smaller size and the hypertriton is even more suppressed as a result of its much larger size.
\end{abstract}

\maketitle

\section{Introduction}
\label{introduction}

Besides the production of the quark-gluon plasma (QGP)~\cite{Shu14,Aki15,Bra16}, relativistic heavy-ion collisions have also led to the production of anti-nuclei~\cite{Abe09,Aga11,Sha11,Ada16} and the discovery of anti-hypernuclei~\cite{Abe10,Ada16-1}.   More recently,  light nuclei production in relativistic heavy-ion collisions have further been used to search for the possible critical point~\cite{Ste06,Fuk11,Luo17,Yin:2018ejt} in the phase diagram of strongly interacting quark matter~\cite{KJSUN17,KJSUN18,Shuryak:2018lgd,Yu:2018kvh}.  However, how and when these light nuclei are produced during relativistic heavy-ion collisions are still under debate   because of their small binding energies and finite sizes~\cite{Oh:2009gx,Fec16,Mrowczynski:2016xqm,Xu:2018jff,Baz18,Braun-Munzinger:2018hat,Oliinychenko:2018ugs,Bellini:2018epz,Bugaev:2018klr}.  On the one hand, they are assumed to be produced at hadronization of the  QGP created in these collisions as in the statistical model for particle production~\cite{And11,And16}. On the other hand, they are described by the coalescence of nucleons and lambda hyperons at the kinetic freeze-out of heavy-ion collisions when the temperature and density of the hadronic matter are low~\cite{Sato:1981ez,Scheibl:1998tk,Sun:2015jta,Sun:2015ulc,Zhu:2015voa,Yin:2017qhg,Zhao:2018lyf,Blum:2019suo}.

In recent measurements by the ALICE Collaboration at the LHC, the yield ratios d/p and $^3$He/p from p+p, p+Pb and Pb+Pb collisions at center-of-mass  energies ranging from 900 GeV to 7 TeV have been measured, and they are found to decrease monotonically with decreasing charged particle multiplicity in the collisions~\cite{Ada16,Ant16,Ach18,Ach18-1}. In particular, the ratio d/p is suppressed by more than a factor of  2 in p+p collisions at $\sqrt{s_{NN}}= 7$ TeV~\cite{Ach18} compared to that in central Pb+Pb collisions at $\sqrt{s_{NN}}= 2.76$ TeV~\cite{Ada16}.  Because of the high collision energies, the produced matter  consists of nearly equal number of particles and antiparticles, and it reaches almost the same temperature  of $T_c\approx 154$  MeV~\cite{Aok06,Aok09,Baz12,Baz14}  at which the initially produced QGP is transformed to the hadronic matter~\cite{Bra07}. Therefore,   almost the same chemical freeze-out temperature occurs in p+p, p+Pb and Pb+Pb collisions~\cite{Sta13,Sha18}, and the only difference between these colliding systems is the size of produced matter or the total number of produced particles.  This is confirmed by the two-pion  correlation measurements through the Hanbury-Brown Twiss (HBT) interferometry, which gives the Gaussian source radii in p+p and Pb+Pb collisions  that are about 2 fm and 10 fm, respectively~\cite{Ada15-1,Abelev:2014pja}. 

In the statistical hadronization approach  based on the grand canonical ensemble, all hadrons produced in heavy-ion collisions at the LHC energies are in thermal and chemical equilibrium, and their yield ratios are determined only by  the chemical freeze-out temperature, as the baryon chemical potential is nearly zero in collisions at such high energies~\cite{Sta13}.  For instance, the measured proton to pion ratio  is about 5$\times10^{-2}$ in  p+p, p+Pb and Pb+Pb collisions, which is consistent with the statistical model prediction based on the grand canonical ensemble.   For the yield ratios of d/p and $^3$He/p, the predicted respective values of about 3.6$\times10^{-3}$ and $1.0\times10^{-5}$ for central Pb+Pb collisions at $\sqrt{s_{NN}}=2.76$ TeV from the statistical model are also in nice agreement with the experimental data. These values are, however,  much larger than those from p+p collisions  at the LHC~\cite{Ach18}. To explain the suppressed production of light nuclei in  collisions of such a small system, the statistical model has been modified to use the canonical ensemble to take into account the conservation of baryon  number, electric charge and the strangeness~\cite{Vov18}. The resulting ratios of light nuclei to proton in these collisions are, however, too small compared with the experimental data unless the canonical correlation volume for exact charge conservations is taken to span three  units of rapidity, instead of the usual one unit of rapidity for collisions with large particle multiplicity, or using a higher chemical freeze-out temperature of 170 MeV than the usual value of 155 MeV for collisions of large systems. 

In the coalescence model, the formation probability of a light nucleus in a heavy-ion collision depends not only on the thermal properties and volume of the nucleon and hyperon emission source but also on the internal wave function of the light nucleus.  The small size of the emission source in p+p collisions is expected to significantly reduce the phase-space volume in which a light nucleus can be formed, leading to a suppression of  its production. Using a schematic coalescence model based on nucleons from the UrQMD model by allowing a deuteron to be formed from a pair of proton and neutron when their separation in phase-space is less than certain value, it is found in Ref.~\cite{Som18} that this model can give a good description of the experimental data on the d/p ratio in p+p, p+A, and A+A collisions at the LHC.

In this Letter, we use a more realistic coalescence model to study the system size or charged particle multiplicity dependence of the d/p and $^3$He/p ratios by taking into account the finite size of deuteron and $^3$He through their internal wave functions.  Our  results on these yield ratios are found in good agreement with available experimental data. We also confirm that the $^3$He/p ratio has a stronger system size dependence than the d/p ratio as  helium-3 has three nucleons and is thus more sensitive to the spatial distribution of nucleons in the emission source. For the triton $^3$H, we find that its production is 10\%-30\% larger  than that of helium-3 and thus less suppressed in p+p collisions because of its smaller matter radius. For the hypertriton $^3_\Lambda$H,  the  $^3_\Lambda$H/$\Lambda$ ratio in collisions with small charged particle multiplicity is found, on the other hand, much more suppressed than the $^3$He/p ratio, and the suppression further depends on whether  the $^3_\Lambda$H is  produced from the coalescence of n-p-$\Lambda$ or d-$\Lambda$.

\section{ Light nuclei production in coalescence model}

Although the coalescence model has been used in various ways for studying light nuclei production in nuclear reactions~\cite{But63,Sch63,Bond:1977fd,Kapusta:1980zz,Sat81,Gyulassy:1982pe,Baltz:1993jh,Kahana:1996bw,Mattiello:1996gq}, we follow in the present study that employed in Refs.~\cite{KJSUN17,KJSUN18}. In this approach,  the formation probability of a light nucleus in heavy-ion collisions is given by the overlap of the  nucleon phase-space distribution functions in the emission source with the Wigner function of the light nucleus, which is obtained from the Wigner transform of its internal wave function~\cite{Chen:2003qj,Chen:2003ava}. For deuteron production in heavy-ion collisions, its number from the coalescence model is given by
\begin{eqnarray}
N_\text{d}&=&g_\text{d}\int \text{d}^3{\bf x}_1\int \text{d}^3{\bf k}_1\int \text{d}^3{\bf x}_2\int \text{d}^3{\bf k}_2 f_n({\bf x}_1,{\bf k}_1) \notag \\
&&f_p({\bf x}_2,{\bf k}_2)W_\text{d}({\bf x}_1-{\bf x}_2,({\bf k}_1-{\bf k}_2)/2), \label{Eq1}
\end{eqnarray}
where $g_{\rm d}=3/4$ is the statistical factor for forming a spin one deuteron from spin half proton and neutron~\cite{Sat81,Zhao:2018lyf}, $f_{p,n}({\bf x},{\bf k})$ are the neutron and proton phase-space distributions, and $W_{\rm d}({\bf x},{\bf k})$ is the Wigner function of the deuteron. 

Since the nucleon coalescence is a local process, one can neglect the effect of collective flow on nucleons and take their phase-space distributions in a thermalized expanding spherical fireball of kinetic freeze-out temperature $T_K$ and  radius $R$ to be 
\begin{eqnarray}
f_{p,n}({\bf x},{\bf k})=\frac{N_{p,n}}{(2\pi)^3(mT_K R^2)^{\frac{3}{2}}}~e^{-\frac{k^2}{2mT_K}-\frac{x^2}{2R^2}},
\end{eqnarray}
with $m$ being the nucleon mass, and they are normalized to their numbers $N_{p,n}$ according to $N_{p,n}=\int \text{d}^3{\bf x}\int \text{d}^3{\bf k}f_{p,n}({\bf x},{\bf k})$.

Using the harmonic oscillator wave function for the internal wave function of the deuteron, which is usually assumed in the coalescence model for deuteron production, its Wigner function then has the Gaussian form~\cite{Sun:2015jta,Sun:2015ulc,Zhu:2015voa}, 
\begin{eqnarray}
W_\text{d}({\bf x},{\bf k})=8~e^{-\frac{x^2}{\sigma^2}}~e^{-\sigma^2 k^2}, \label{Eq2}
\end{eqnarray}
with the normalization $\int \text{d}^3{\bf x}\int \text{d}^3{\bf k}~W_d({\bf x},{\bf k})=(2\pi)^3$.  Transforming the proton and neutron coordinates ${\bf x}_1$ and ${\bf x}_2$ as well as their momenta ${\bf k}_1$ and ${\bf k}_2$  to their center-of-mass reference frame, 
\begin{eqnarray}
{\bf X}&=&\frac{{\bf x}_1+{\bf x}_2}{2},\quad {\bf x}={\bf x}_1-{\bf x}_2, \notag \\
{\bf K}&=&{\bf k}_1+{\bf k}_2,\quad {\bf k}=\frac{{\bf k}_1-{\bf k}_2}{2},  \label{Eq3}
\end{eqnarray}
the integrals in Eq.~(\ref{Eq1}) can then be straightforwardly evaluated, leading to
\begin{eqnarray}
N_\text{d}&=&\frac{8g_{\rm d} N_pN_n}{(2\pi)^6(mT_KR^2)^{3}}  \int \text{d}^3{\bf X}e^{-\frac{X^2}{R^2}}\int \text{d}^3{\bf x}~e^{-(\frac{1}{\sigma^2}+\frac{1}{4R^2})x^2}\notag \\
&&\int \text{d}^3{\bf K}~e^{-\frac{K^2}{4mT_K}}\int \text{d}^3{\bf k}~e^{-k^2(\sigma^2+\frac{1}{mT_K})}\nonumber\\
&=&\frac{3N_nN_p}{4(mT_K R^2)^{3/2}}\frac{1}{\left(1+\frac{1}{mT_K\sigma^2}\right)^{3/2}}\frac{1}{(1+\frac{\sigma^2}{4R^2})^{3/2}}.  \label{Eq4}
\end{eqnarray}

The parameter $\sigma$ in Eq.(\ref{Eq2}) is related to the root-mean-square matter radius $r_{\text d}=1.96~{\rm fm}$ of deuteron~\cite{Rop09} by $\sigma  = \sqrt{8/3}~r_{\text d}\approx 3.2$ fm.  For the kinetic freeze-out temperature $T_K$ of nucleons, it is typically of the order of $100$ MeV. We therefore have $mT_K\gg 1/\sigma^2$, and the yield ratio d/p is then approximately given by
\begin{eqnarray}
\frac{N_\text{d}}{N_p} \approx \frac{3N_n}{4(mT_KR^2)^{3/2}}\frac{1}{\left[1+(\frac{1.6~\text{fm}}{R})^2\right]^{3/2}.}  \label{Eq5}
\end{eqnarray}
The last factor in the above equation describes the suppression of deuteron production due to its finite size relative to that of the nucleon emission source. Its value approaches unity as the source radius $R$ becomes much larger than the size of deuteron, while it is significantly smaller than  unity when $R$ is close to or less than 1.6 fm.  The  factor $C_1=\frac{3N_n}{4(mT_KR^2)^{3/2}}$ in Eq.~(\ref{Eq5}) corresponds to the d/p ratio in the limit of large nucleon emission source when the suppression effect due to finite deuteron size is negligible, and it is directly related to the entropy per nucleon in a nuclear collision, which remains essentially unchanged after chemical freeze-out~\cite{Sie79}. Therefore, the value of $C_1$ is expected to be similar in p+p, p+Pb and Pb+Pb collisions at the LHC. From the d/p ratio measured in central Pb+Pb collisions at $\sqrt{s_{NN}}= 2.76$ TeV, a value of about  $4.0 \times 10^{-3}$ is obtained from Eq.(\ref{Eq5}) for $C_1$. Using this value,  Eq.~(\ref{Eq5}) can be rewritten as
\begin{eqnarray}
\frac{N_\text{d}}{N_p}\approx \frac{4.0\times10^{-3}}{\left[1+(\frac{1.6~\text{fm}}{R})^2\right]^{3/2}}, \label{Eq5-1}
\end{eqnarray}
where the value of $R$ can be calculated from 
\begin{eqnarray}
R=\frac{(3N_n)^{1/3}}{[4C_1(mT_K)^{3/2}]^{1/3}}. \label{Eq5-1-1}
\end{eqnarray}
using the neutron number $N_n$, which is the same as the proton number in collisions at the LHC energies because of the vanishing isospin chemical potential, and the kinetic freeze-out temperature $T_K$  extracted from  measured charged particle spectra. 

\begin{figure}[!h]
\includegraphics[scale=0.46]{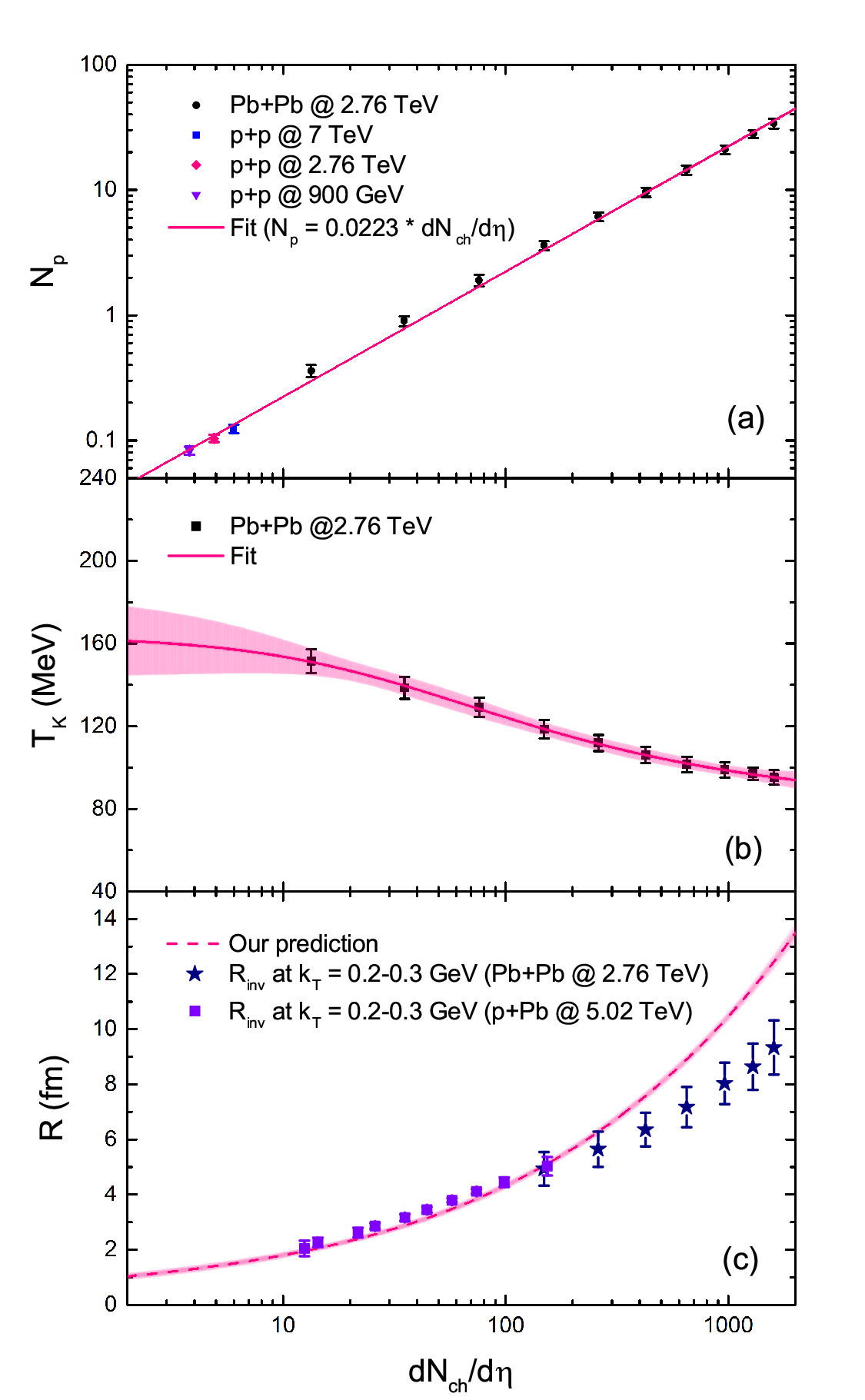}
\caption{Charged particle multiplicity dependence of the proton number $N_p$ (panel (a)), kinetic freeze-out temperature $T_K$ (panel (b)) and the radius of  emission source $R$ (panel (c)).  
~\cite{Abe13,Ada15-2,Ada15-1}.
The solid line in panel (a) represents a linear fit to the data. The solid line in panel (b) is the fit to the data  using Eq.(\ref{Eq5-0}) with the shadow region surrounding the line denoting the uncertainties.  The uncertainties of the data points shown in panel (b) are only statistical~\cite{Abe13}. The dashed line in panel (c) is the  predicted radius of emission source with uncertainties given by the shaded band.  Except the experimental data shown by solid squares with error bars in panel (c), which are from the ATLAS Collaboration~\cite{Aaboud:2017xpw}, all other experimental data are from the ALICE Collaboration ~\cite{Abe13,Ada15-2,Ada15-1}.}
\label{Fig0}
\end{figure}

Shown in panel (a) of Fig.~\ref{Fig0} by symbols with error bars
is the charged particle multiplicity dependence of the proton number measured by the ALICE Collaboration~\cite{Abe13,Ada15-2}.  The dependence is seen to be essentially linear  and can be well parametrized by $N_p = 0.0223\times \text{d}N_\text{ch}/d\eta$ shown by the solid line. 

Panel (b) of Fig.~\ref{Fig0}  shows the charged particle multiplicity dependence of the kinetic freeze-out temperature. The solid circles with error bars are from  the ALICE Collaboration based on a blast wave model fit to the experimental data~\cite{Abe13}. It is seen that $T_K$ increases as $\text{d}N_\text{ch}/d\eta$ decreases, and it can be fitted by the function 
\begin{eqnarray}
T_K = T_0+T_1\left[1+(q-1)\times\frac{\text{d}N_\text{ch}/d\eta}{M}\right]^{-\frac{1}{q-1}}, \label{Eq5-0}
\end{eqnarray}
in terms of the four parameters $T_0=80.6\pm 31.0$ MeV, $T_1=83.0\pm 46.9$ MeV, $M=67.3\pm76.3$, and $q=3.33\pm3.25$ after taking into account the errors in the extracted $T_K$. The corresponding uncertainty of $T_K$ at any charged particle multiplicity can be obtained from $\Delta T_K = \left[2\sum_{i,j}\frac{\partial T_K}{\partial x_i}(H^{-1})_{ij}\frac{\partial T_K}{\partial x_j}\right]^{1/2}$, where $x_i$ is one of the four parameters in Eq.(\ref{Eq5-0}). The  Hessian matrix $H$  in our chi-square fit to the empirically extracted $T_K$ is given by  
\begin{equation}       
H = \left(                 
  \begin{array}{cccc}  
    1.27 & 4.33\times 10^{-1} & 1.80\times 10^{-1} &9.33\\ 
    4.33\times 10^{-1} & 1.93\times 10^{-1} & 6.90\times 10^{-2} &2.71\\  
    1.80\times 10^{-1} & 6.90\times 10^{-2} & 2.79\times 10^{-2} &1.26\\  
    9.33& 2.71 & 1.26 &7.44\times 10^{1}\\  
  \end{array}
\right).         
\end{equation}
The  fitted charged particle multiplicity dependence of $T_K$ is shown  in panel (b) of Fig.~\ref{Fig0} by the solid line with the shaded band denoting its uncertainties. We note that the function in Eq.(\ref{Eq5-0}) has a similar form as the Tsallis distribution for the single particle energies in a non-extensive system~\cite{Tsallis:1987eu}.

With the  above determined values of $N_n$ and $T_K$, we can evaluate from Eq.~(\ref{Eq5-1-1}) the charged particle multiplicity dependence of the radius $R$ of  emission source. The result is depicted  in panel (c) of Fig.~\ref{Fig0} by the dashed line with the theoretical uncertainties given by the shaded band, which turns out to be quite small.  Also shown in this panel by solid squares and stars with error bars are the one-dimensional femtoscopic radius $R_\text{inv}$ of the Gaussian emission source extracted, respectively, by the ATLAS Collaboration~\cite{Aaboud:2017xpw} and by the ALICE Collaboration~\cite{Ada15-1} from the two-pion interferometry measurements~\cite{Lis05}  for pion pairs of transverse momentum $k_T=$ 0.2-0.3 GeV. It is seen that the predicted $R$ is larger than $R_\text{inv}$ for central Pb+Pb collisions. This  is likely due to the large radial flow in central Pb+Pb collisions, which  would lead to a smaller apparent Gaussian radii of an emission source.

\begin{figure}[h]
\includegraphics[scale=0.46]{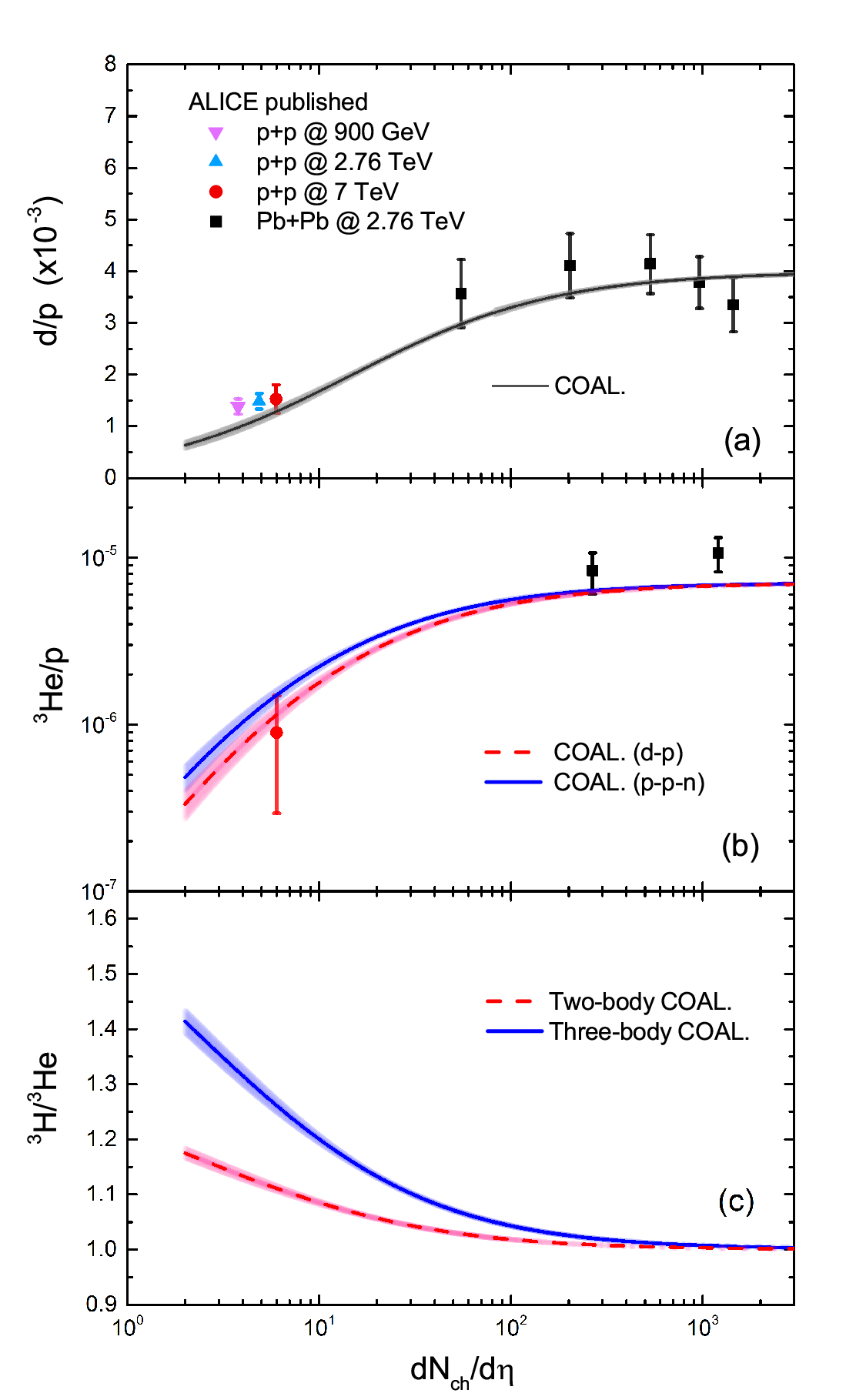}
\caption{Charged particle multiplicity dependence of the yield ratios d/p, $^3$He/p and $^3$H/$^3$He. The lines denote the predictions of coalescence model with theoretical uncertainties on the emission source radius given by the  shaded band.  Experimental data from the ALICE Collaboration are shown by symbols with error bars~\cite{Abe13,Ada15-2,Ada16,Ach18}. }
\label{Fig1}
\end{figure}

With the information on the radius of nucleon emission source, the d/p ratio can then be calculated from Eq.~(\ref{Eq5-1}), and its dependence on the charged particle multiplicity is shown in panel (a) of Fig. \ref{Fig1}. Compared with the measured ratio in central Pb+Pb collisions at $\sqrt{s_{NN}}= 2.76$ TeV~\cite{Abe13,Ada16} and in p+p collisions at $\sqrt{s_{NN}}=$~900 GeV, 2.76 TeV and 7 TeV~\cite{Ada15-2,Ach18}, the theoretical results are in nice agreement with the data for all charged particle multiplicities.  Our results are consistent with those from a schematic coalescence model based on kinetic freeze-out nucleons from the UrQMD model~\cite{Som18}. We note that the finite deuteron size   suppresses not only the total yield ratio of deuteron to proton as studied here  but also  their ratio as a function of  transverse momentum~\cite{Bellini:2018epz}. We also note that the size effect in the coalescence model was firstly studied in Refs.~\cite{Hagedorn:1960zz,Hagedorn:1962mot} at early 1960s and then comprehensively studied  in Refs.~\cite{Scheibl:1998tk}. 

Similarly, we can calculate the charged particle multiplicity dependence of the $^3$He/p ratio by extending the formalism for deuteron production from proton and neutron coalescence to the production of helium-3 from  the coalescence of two protons and one neutron as in Refs.~\cite{Sun:2015jta,Sun:2015ulc,Zhu:2015voa,Yin:2017qhg,Zhao:2018lyf}.  The resulting yield ratio $^3$He/p is given by
\begin{eqnarray}
\frac{N_{^3\text{He}}}{N_p} \approx  \frac{N_nN_p}{4(mT_KR^2)^{3}}
\frac{1}{\left(1+\frac{r_{^3\text{He}}^2}{2R^2}\right)^{3}},
\label{Eq7}
\end{eqnarray}
where  $r_{^3\text{He}}=1.76$ fm is the matter radius of helium-3~\cite{Rop09}. In obtaining the above equation, we have included the statistical factor of 1/4 for forming a spin 1/2 helium-3 from three spin 1/2 nucleons and used the condition $mT_K\gg 1/r_{^3\text{He}}^2$.  With the factor $C_2=\frac{N_nN_p}{4(mT_KR^2)^{3}}= 4C_1^2/9=7.1\times 10^{-6}$, determined from the value of $C_1$, Eq.(\ref{Eq7}) becomes
\begin{eqnarray}
\frac{N_{^3\text{He}}}{N_p}\approx \frac{7.1\times 10^{-6}}{\left[1+(\frac{1.24~\text{fm}}{R})^2\right]^{3}}.
\label{Eq8}
\end{eqnarray}

We also consider $^3$He production from the coalescence of a deuteron and a proton. In this case, the root-mean-square radius of $^3$He can be estimated as $r_{^3\text{He}}\approx(3/8)^{1/2}\sqrt{\langle r_{\text{pd}} \rangle^2}=1.15$ fm with $\sqrt{\langle r_{\text{pd}} \rangle^2}\approx2.6$ fm being the distance between proton and the center of mass of the deuteron inside the helium-3.  Using the statistical factor of 1/3 for the coalescence of a spin 1 deuteron and a spin 1/2 proton to $^3$He, the $^3$He/p ratio  is then 
\begin{eqnarray}
\frac{N_{^3\text{He}}}{N_p}\approx \frac{7.1\times 10^{-6}}{\left[1+(\frac{1.15~\text{fm}}{R})^2\right]^{3/2}\left[1+(\frac{1.6~\text{fm}}{R})^2\right]^{3/2}},
\label{Eq9}
\end{eqnarray}
where the suppression factor for deuteron production has been included.  As shown in panel (b) of Fig. \ref{Fig1},  the contribution from the coalescence of deuteron and proton is larger than that from the coalescence of two protons and one neutron in collisions of small charged particle multiplicities, although the two  processes give similar contributions to $^3$He production in collisions of large charged particle multiplicities. Besides, the theoretical results are found in nice agreement with the data at $\text{d}N_\text{ch}/d\eta < 1000$, while they are slightly smaller than the data in the most central Pb+Pb collisions at $\sqrt{s_{NN}}= 2.76$ TeV. 

The above calculation for helium-3 production can be straightfowardly extended to triton ($^3$H) production. Because of its  smaller radius of $r_{^3\text{H}}=~1.59$ fm~\cite{Rop09} than helium-3,  triton production in collisions with low multiplicities is expected to be less suppressed  than helium-3.   For instance, the  $^3$H/$^3$He ratio  is 
\begin{eqnarray}
\frac{N_{^3\text{H}}}{N_{^3\text{He}}}\approx \frac{\left[1+(\frac{1.24~\text{fm}}{R})^2\right]^{3}}{\left[1+(\frac{1.12~\text{fm}}{R})^2\right]^{3}}
\label{Eq9-1}
\end{eqnarray}
from the three-body coalescence and 
\begin{eqnarray}
\frac{N_{^3\text{H}}}{N_{^3\text{He}}}\approx \frac{\left[1+(\frac{1.15~\text{fm}}{R})^2\right]^{3/2}}{\left[1+(\frac{1.039~\text{fm}}{R})^2\right]^{3/2}}
\label{Eq9-2}
\end{eqnarray}
from the two-body coalescence.  Shown in panel (c) of Fig. 1 is the  $^3$H/$^3$He yield ratio as a function of charged particle multiplicity.  It is seen that this ratio indeed increases with decreasing charged particle multiplicity, particularly for triton and helium-3 production from three-body coalescence. For instance, this ratio in p+p collisions at $\text{d}N_\text{ch}/d\eta = 5$  is predicted to be 1.1 if triton and helium-3 are produced from two-body coalescence but increases to 1.3  if they are produced from three-body coalescence, suggesting a 10\%-30\% enhancement in the production  of triton than heilium-3 in p+p collisions. Future measurements of  the triton yield in p+p collisions can be used to testify this result. 

\section{$^3_\Lambda$H production in coalescence model}

To study the production of $^3_{\Lambda}$H in collisions of small systems, we first note that  $^3_{\Lambda}$H is the lightest known nucleus with strangeness, and it has a small binding energy of only B$_\Lambda$= 2.35 MeV and a large root-mean-square radius of $r_{^3_\Lambda\text{H}}\approx4.9$ fm~\cite{Nem00}.   Besides being a bound state of proton, neutron and $\Lambda$-hyperon, the hypertriton can also be considered as a bound state of a deuteron and a $\Lambda$-hyperon with a binding energy B$_\Lambda$ = 0.13 $\pm$ 0.05 MeV \cite{Jur73} and a distance of $r_{\Lambda\text{d}}\approx 10$ fm~\cite{Nem00} between deuteron and  $\Lambda$-hyperon.  Because of its large size, the production of $^3_\Lambda$H in collisions of small systems is expected to be much more suppressed than that of helium-3. We note that the study of $^3_\Lambda$H production in relativistic heavy-ion collisions including both the coalescence of p-n-$\Lambda$ and of d-$\Lambda$ has recently been reported in~Ref.~\cite{Zha18}.  According to this study, the process of p-n-$\Lambda$ coalescence is more important than that of the d-$\Lambda$ coalescence for hypertrion production, and the hypertriton yield in relativistic heavy-ion collisions is essentially determined at the time when nucleons and deuterons freeze out, although it still undergoes reactions with pions.

Similar to  helium-3 production, the yield ratio $^3_\Lambda\text{H}/\Lambda$ is given by
\begin{eqnarray}
\frac{N_{^3_\Lambda\text{H}}}{N_\Lambda} \approx \frac{7.1\times 10^{-6}}{\left[1+(\frac{3.46~\text{fm}}{R})^2\right]^{3}}
\label{Eq10}
\end{eqnarray}
for hypertriton production from the coalescence of proton, neutron and $\Lambda$-hyperon, and
\begin{eqnarray}
\frac{N_{^3_\Lambda\text{H}}}{N_\Lambda}  \approx \frac{7.1\times 10^{-6}}{\left[1+(\frac{4.2~\text{fm}}{R})^2)^{3/2}(1+(\frac{1.6~\text{fm}}{R})^2\right]^{3/2}}.
\label{Eq11}
\end{eqnarray}
for hypertriton production from the coalescence of d and $\Lambda$. In obtaining Eq.(\ref{Eq10}) for the three-body coalescence process, we have taken the root-mean-square radius of $^3_\Lambda$H as $r_{^3_\Lambda\text{H}}\approx(3/8)^{1/2}\sqrt{\langle r_{\Lambda\text{d}} \rangle^2}=4.2$ fm. Also, we have neglected the mass difference of the constituent particles in obtaining above expressions since its effect is small.

\begin{figure}
\includegraphics[scale=0.34]{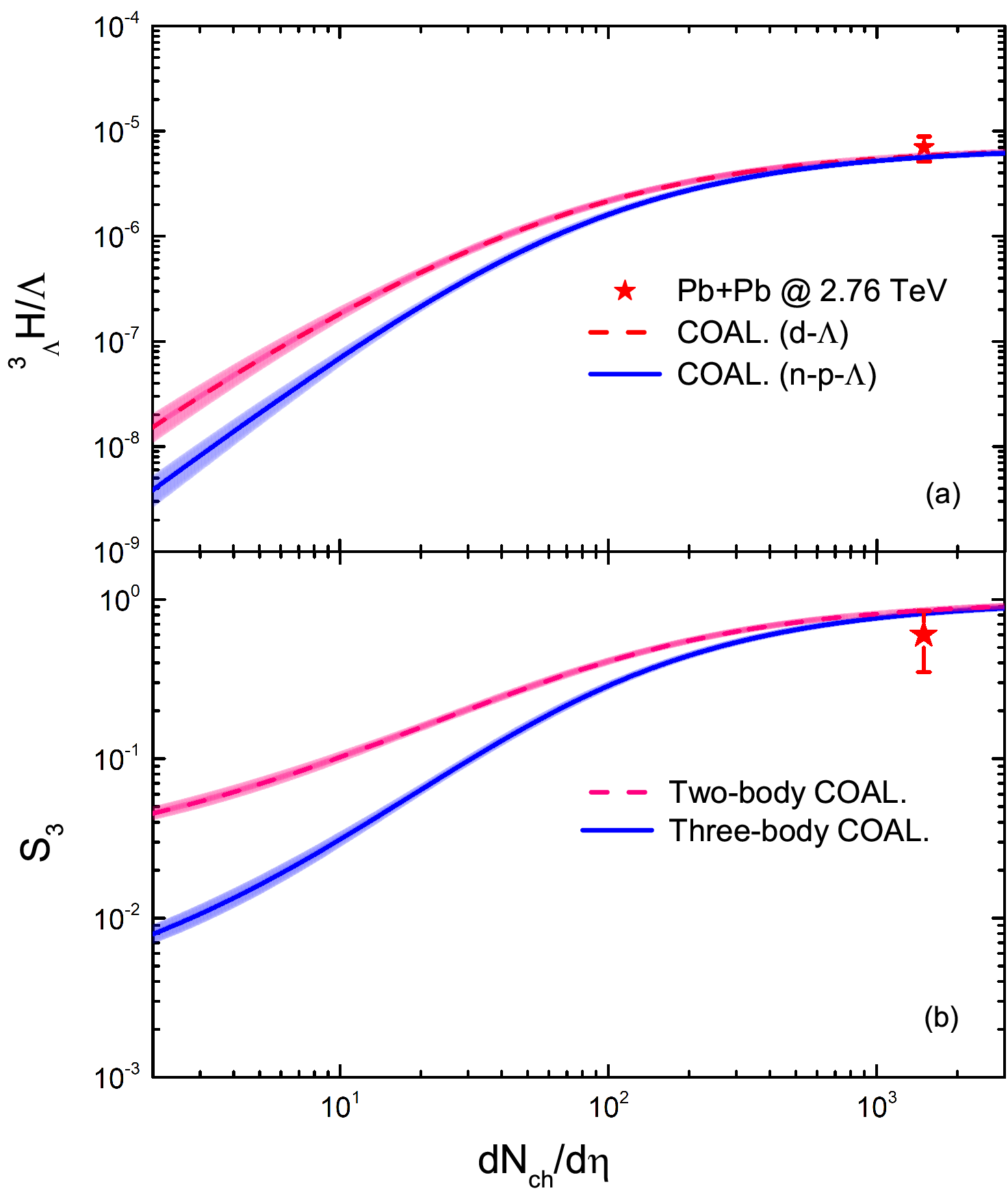}
\caption{\protect Charged particle multiplicity dependence of the yield ratio $^3_\Lambda$H/$\Lambda$ and the $S_3$ factor.  Predictions from the coalescence model are shown by solid lines for the three-body coalescence and dashed lines for the two-body coalescence with theoretical uncertainties given by  shaded bands. Experimental data from the ALICE Collaboration~\cite{Ada16,Ada16-1} are shown by solid stars with error bars.}
\label{Fig2}
\end{figure}

In panel (a) of Fig.~\ref{Fig2}, we show the  charged particle multiplicity dependence of the yield ratio $^3_\Lambda\text{H}/\Lambda$ in Pb+Pb collisions at $\sqrt{s_{NN}}= 2.76$ TeV. The dashed and solid lines represent results from the two-body and the three-body coalescence, respectively.  No significant difference is seen between these two processes when $\text{d}N_\text{ch}/d\eta>100$, and both agree very well the experimental data shown by the solid star with error bar measured by the ALICE Collaboration for central collisions.  For $\text{d}N_\text{ch}/d\eta\sim 10$, both production processes give a yield ratio $^3_\Lambda\text{H}/\Lambda$ that is two-order of magnitude less than in central Pb+Pb collisions. 

We further investigate the strangeness population factor $S_3$, which is a double ratio defined  by $S_3 = ^3_{\Lambda}$H/($^3$He$\times \Lambda/$p)~\cite{Arm04}. As suggested in Ref.~\cite{Arm04}, the value of $S_3$ should be about one in  the coalescence model for particle production. It was also argued in Ref.~\cite{Zha10} that this factor might be a good signal for studying the local correlation between baryon number and strangeness in a quark-gluon plasma~\cite{Koc05}, providing thus a valuable probe of the onset of deconfinement in relativistic heavy-ion collisions.  The system size dependence of $S_3$ can be calculated from Eqs.~(\ref{Eq8}) and (\ref{Eq10}),  for $^3$He and $^3_\Lambda$H production from three-body coalescence and  from Eqs.~(\ref{Eq9}) and (\ref{Eq11}) for their production from two-body coalescence.  Results for Pb+Pb collisions at $\protect\sqrt{s_{NN}}=2.76$ TeV are shown in panel (b) of Fig.~\ref{Fig2} for both the two-body (dashed line) and the three-body (solid line) coalescence. One can see that  the $S_3$ factor in central collisions is close to unity in both cases, similar to the experimental value shown by the solid star with error bar measured by the ALICE Collabortion ~\cite{ALICE15H}.  Also, there is no significant charged particle multiplicity dependence in the $S_3$ factor given by the two coalescence processes when $\text{d}N_\text{ch}/d\eta>100$.  However, they start to deviate when $\text{d}N_\text{ch}/d\eta$ becomes smaller, with the three-body coalescence giving a much smaller value than the two-body coalescence as a result of the suppressed production of hypertriton from three-body coalescence in small systems.

\section{Conclusions}

In summary, based on the coalescence model in full phase space, we have studied the dependence of deuteron, heilium-3, and triton production in nuclear collisions at energies available from the LHC on the charged particle multiplicity of the collisions. For the nucleon distributions, they are assumed to come from a thermalized hadronic matter at the kinetic freeze-out of heavy-ion collisions with its temperature taken from the empirical fit to measured particle spectra and its size determined by assuming that the entropy per baryon   is independent of the colliding system. We have found that the yield ratios d/p and $^3$He/p are significantly reduced once the charged particle multiplicity is below about 100 as a result of the non-negligible deuteron and $^3$He sizes compared to that of the nucleon emission source. Our results thus provide a natural explanation for the observed suppression of deuteron and $^3$He production in p+p collisions by the ALICE Collaboration at the LHC. They also demonstrate the importance of the internal structure of light  nuclei on their production in collisions of small systems.  We have further found that the production of triton is 10\%-30\% larger than that of helium-3 in p+p collisions because  of its smaller matter radius. This enhancement of  $^3$H/$^3$He ratio can  be  tested in future measurements.

We have also used this model to study the charged particle multiplicity dependence of hypertriton production in Pb+Pb collisions at the LHC by considering both the three-body process of p-n-$\Lambda$ coalescence and the two-body process of d-$\Lambda$ coalescence.   Because of the much larger $^3_\Lambda$H radius than those of deuteron and $^3$He, the yield ratio $^3_\Lambda$H/$\Lambda$ is found to be much more suppressed in collisions with low charged{-}particle multiplicity, particularly for the three-body coalescence process.   We have further studied the charged particle multiplicity dependence of the strangeness population factor $S_3 = ^3_{\Lambda}$H/($^3$He$\times \Lambda/$p), and its value in collisions with small charged particle multipilicity is found to be significantly less than one expected in collisions with large charged particle multiplicity. Future experimental measurements of the yield ratio $^3_\Lambda\text{H}$/$\Lambda$ and the strangeness population factor $S_3$ in collisions of low charged particle multiplicity will be of great interest because it not only can check the prediction of the present study but also provide the possibility to improve our knowledge on the internal structure of $^3_\Lambda$H. 


\begin{acknowledgments}
This work was supported in part by the US Department of Energy under Contract No. DE-SC0015266 and the Welch Foundation under Grant No. A-1358 as well as  by BMBF through the FSP202 (F\"{o}rderkennzeichen 05P15RFCA1).
\end{acknowledgments}

\bibliography{Nuclei190310}

\end{document}